\documentclass[runningheads]{llncs}
\usepackage[T1]{fontenc}
\usepackage{amsmath, xcolor}
\usepackage{graphicx,verbatim}
\usepackage{hyperref}
\begin{document}
\title{Anatomy-constrained modelling of image-derived input functions in dynamic PET using multi-organ segmentation}
\titlerunning{Anatomical modelling of image-derived input functions}

\author{Valentin Langer\inst{1} \and
Kartikay Tehlan\inst{1,2}\orcidID{0000-0003-2417-1202} \and \\
Thomas Wendler\inst{1,2,3,4}\orcidID{0000-0002-0915-0510}}
\authorrunning{V. Langer et al}
\institute{Department of diagnostic and interventional Radiology and Neuroradiology, University Hospital Augsburg, Stenglinstr. 2, 86156 Augsburg, Germany \email{kartikay.tehlan@med.uni-augsburg.de} \and 
Computer-Aided Medical Procedures and Augmented Reality, Technical University of Munich, Boltzmannstr. 3, 85748 Garching bei München, Germany \and
Digital Medicine, University Hospital Augsburg, Gutenbergstr. 7, 86356, Neusäß, Germany \and
Center of Advanced Analytics and Predictive Sciences, University of Augsburg, Universitätsstr. 2, 86159 Augsburg, Germany}

\maketitle              %
\begin{abstract}
Accurate kinetic analysis of [$^{18}$F]FDG distribution in dynamic positron emission tomography (PET) requires anatomically constrained modelling of image-derived input functions (IDIFs). Traditionally, IDIFs are obtained from the aorta, neglecting anatomical variations and complex vascular contributions. This study proposes a multi-organ segmentation-based approach that integrates IDIFs from the aorta, portal vein, pulmonary artery, and ureters. Using high-resolution CT segmentations of the liver, lungs, kidneys, and bladder, we incorporate organ-specific blood supply sources to improve kinetic modelling. Our method was evaluated on dynamic [$^{18}$F]FDG PET data from nine patients, resulting in a mean squared error (MSE) reduction of 13.39\% for the liver and 10.42\% for the lungs. These initial results highlight the potential of multiple IDIFs in improving anatomical modelling and fully leveraging dynamic PET imaging. This approach could facilitate the integration of tracer kinetic modelling into clinical routine.\\
The code is available under: \url{https://github.com/tinolan/curve_fit_multi_idif}


\end{abstract}

\section{Introduction}

The enhanced temporal and spatial sensitivity of long axial field-of-view Positron Emission Tomography (PET) scanners \cite{volpi2023update} has enabled the use of dynamic [$^{18}$F]FDG PET scans for early metastasis detection, differentiation between malignant and inflammatory lesions, and assessment of treatment response \cite{wu2024role,skawran2022can,de2008chemotherapy}. These applications are based on differences in glucose metabolism. For instance, cancer cells primarily upregulate GLUT1 and GLUT3 transporters, as well as HK2, to sustain high glycolytic rates for proliferation and survival \cite{pliszka2021glucose,aft2002evaluation}. In contrast, macrophages mainly rely on GLUT1-mediated glucose uptake \cite{cornwell2023glucose,obaid2021lncrna}. The analysis of time-activity curves (TACs) in FDG PET, using models such as the two-compartment model (2CM), allows differentiation of these metabolic patterns \cite{dimitrakopoulou2021kinetic}.

The kinetic parameters $K_1$ (glucose uptake), $k_2$ (glucose efflux), and $k_3$ (phosphorylation), along with $V_B$ (vascular volume fraction), derived from 2CM, characterize tissue glucose metabolism. These parameters can be visualized as parametric images, enhancing diagnostic capabilities beyond static PET imaging \cite{dimitrakopoulou2021kinetic}. The TAC of a voxel is used to compute parametric images given an input function, $A(t)$ modelling the radioactive concentration in arterial blood.

A common non-invasive approach to obtain $A(t)$ is to use the radioactivity concentration in the descending aorta \cite{volpi2023update} as the image-derived input function (IDIF). However, this approach assumes direct arterial supply to tissues, overlooking complex vascular pathways that influence tracer distribution. Although IDIFs can also be derived from smaller arteries \cite{tang2025assessment}, several organs receive tracer input from alternative sources, such as the portal vein, pulmonary arteries, and ureters \cite{brix2001quantification}.

Kinetic parameter estimation from TACs and IDIFs is typically achieved through non-linear regression methods, minimizing discrepancies between measured and estimated TACs. Given the significance of parametric images for differential diagnosis in FDG PET \cite{wumener2022dynamic}, replacing the assumption of uniform aortic blood tracer concentration with an anatomically accurate model is worth investigating. 

We propose leveraging multi-organ segmentations from co-registered high-resolution CT scans to incorporate anatomically correct blood circulation sources for kinetic modelling. Specifically, our contributions include:

\begin{itemize}
    \item an organ-specific IDIF model for the liver, lungs, kidneys, and bladder,
    \item a method to extract relevant vasculature and organs for calculating multiple IDIFs, and
    \item the evaluation of the approach on an FDG dynamic PET dataset.
\end{itemize}

\section{Related Works}

Tracer kinetics in organs like the liver, lungs, and kidneys differ from other tissues, and anatomically accurate compartment models significantly increase the number of unknown parameters, making parameter estimation more complex \cite{sommariva2021mathematical}. Animal studies highlight the importance of dual input from the hepatic artery and portal vein for estimating tracer metabolism, including FDG, in porcine livers \cite{winterdahl2011tracer}. Notably, arterial blood sampling combined with proper modelling can provide reasonable approximations. Similar findings by Brix et al. confirm the superiority of dual-input models for improved data fitting \cite{brix2001quantification}.

In human studies, dual-input kinetic parameters (hepatic artery and portal vein for the liver, bronchial and pulmonary arteries for the lungs) have been shown to aid pathological differentiation \cite{zuo2019structural,wang2018dynamic,wang2024high}. These studies derive liver dual-input models from IDIFs using volumes of interest (VOIs) in the descending aorta and optimize the hepatic artery’s contribution to liver blood flow \cite{zuo2019structural,wang2018dynamic}. Similarly, lung tracer supply is modeled using VOIs in the left and right ventricles to extract bronchial and pulmonary artery IDIFs \cite{wang2024high}. Tissue-specific IDIF extraction has been explored as a precursor to voxel-based kinetic parameter estimation \cite{wang2021total}, but high-resolution multi-organ CT segmentation of multiple vasculature and ureters has not yet been proposed.

\section{Methodology}

\subsection{Dataset}
For our analysis, we evaluated 24 patients who underwent dynamic $^{18}$F-FDG PET imaging (65 minutes, acquisition sequence: 2 × 10s, 30 × 2s, 4 × 10s, 8 × 30s, 4 × 60s, 5 × 120s, and 9 × 300s). The in-plane spatial resolution was 3.27 mm full-width at half-maximum (FWHM), with a TOF resolution of 228 ps at 5.3 kBq/mL and a mean injected activity of 235 ± 51 MBq. Imaging was followed by a high-resolution CT scan (voltage: 120 kV, tube current: 25 mA, voxel size: 1.52 × 1.52 × 1.65 mm³) acquired on the same device (Biograph Vision Quadra, Siemens Healthineers, Knoxville, TN, USA).

To ensure physiologically consistent modelling of $A(t)$ and to avoid motion-induced errors, we selected only patients in whom the temporal evolution of tracer concentration in the pulmonary artery, aorta, portal vein, and ureter exhibited a clear sequential increase, with no detectable motion in coronal maximum intensity projection sequences evaluated visually. After applying these criteria, 9 patients remained for analysis.

Due to data protection constraints, we cannot make all data available, yet organ TACs and IDIFs can be found here: 
\begin{center}
    \url{https://doi.org/10.6084/m9.figshare.28509938}
\end{center}

\subsection{Multi-organ Segmentation}

TotalSegmentator \cite{totalsegmentator} (TS) was used to segment the liver, lungs, kidneys, portal vein, pulmonary artery, and urinary bladder. Since TS identifies the portal and splenic veins as a single structure, we manually retained only the segment closest to the liver to ensure accurate localization of the portal vein.

As the ureters are not included in TS, we developed a custom segmentation pipeline based on their spatial relationship with the kidneys. Because the ureters are often challenging to segment directly, the renal pelvis was segmented instead and used as a surrogate in subsequent analyses. First, kidney masks were identified, and their center of mass was computed. An ellipsoid mask was then generated around each kidney, encompassing a significant portion of the expected renal pelvis location. To isolate the renal pelvis, kidney voxels, which include only kidney parenchyma, were removed from the mask, and a connected component analysis was applied slice-wise to retain only the largest component closest to the kidney’s center of mass. This approach resulted in a feasible mask for all 9 patients analyzed.

\subsection{Kinetic Modelling}

The 2CM used for FDG kinetic analysis consists of unmetabolized FDG in the blood plasma (free tracer, $F(\vec{x},t)$) and metabolized FDG trapped inside the cells (bound tracer, $B(\vec{x},t)$). The exchange of FDG between plasma and tissues can be described by the following ordinary differential equations (ODEs) \cite{watabe2017compartmental}:

\begin{equation}
\begin{aligned}
\frac{dF(\vec{x},t)}{dt} &= K_1(\vec{x}) A(t) - (k_2(\vec{x})+k_3(\vec{x}))F(\vec{x},t)\\
\frac{dB(\vec{x},t)}{dt} &= k_3(\vec{x})F(\vec{x},t)\\
C(\vec{x},t) &= F(\vec{x},t) + B(\vec{x},t)
\end{aligned}
\end{equation}

where $A(t)$ is the input function. The solution to these ODEs is given by:

\begin{equation}
C(\vec{x},t) = \frac{K_1(\vec{x})}{k_2(\vec{x})+k_3(\vec{x})} \left[ k_3(\vec{x})+k_2(\vec{x})e^{-(k_2(\vec{x})+k_3(\vec{x}))t} \right] \ast A(t).
\end{equation}

To account for partial blood volume effects, incorporating the vascular fraction $V_B$, the solution is modified as follows:

\begin{equation}
C'(\vec{x},t) = V_B(\vec{x})A(t) + (1-V_B(\vec{x})) C(\vec{x},t).
\end{equation}

To model different tracer input functions, we calculate $A(t)$ as the weighted sum of four IDIFs, namely, the aorta, the portal vein (pv), the pulmonary artery (pa), and the ureters (with respective weights $\alpha$, $\beta$, $\gamma$, and $\delta$). Zhe IDIF was defined as the average PET signal within the CT-based mask of the respective organ:

\begin{equation}
A(t) = \alpha \times \text{IDIF}_{\text{aorta}}(t) + \beta \times \text{IDIF}_{\text{pv}}(t) + \gamma \times \text{IDIF}_{\text{pa}}(t) + \delta \times \text{IDIF}_{\text{ureter}}(t)
\end{equation}

\subsection{Curve fitting}

To estimate the kinetic parameters, we fit the predicted TACs to the measured TACs using non-linear least squares optimization with scipy.optimize.curve\_fit and the mean square error (MSE) as loss. Since we impose physiologically plausible bounds on the parameters, curve\_fit employs the Trust-Region Reflective (TRF) algorithm, which is well-suited for constrained optimization problems. \cite{trfalgorithm}

To ensure biologically meaningful parameter estimation, we constrain $K_1$ and $k_2$ between 0.01 and 10, while $k_3$ is limited to the range 0.001 to 1 \cite{feng2021total,sari_first_2022}. The blood volume fraction $V_b$ is generally restricted between 0.001 and 1; yet, to account for organ-specific characteristics, we further refine its upper limit to 0.15 for the lungs and 0.25 for the liver and kidneys  \cite{eipel2010regulation,pienn2018healthy,effros1967vascular}.

Additional weighting parameters $\alpha$, $\beta$, $\gamma$, and $\delta$ were constrained between 0 and 1, and initialized at 0 are used. For detailed analysis at a per organ-level, we set $\gamma=\delta=0$ for the liver, $\beta=\delta=0$ for the lungs and $\beta=\gamma=0$ for the kidneys. For the kidneys, we also let $\delta$ move between -1 and 1.

\section{Results}

\subsection{Detailed quantitative analysis per organ}

\begin{figure}
    \centering
    \includegraphics[width=0.4\linewidth]{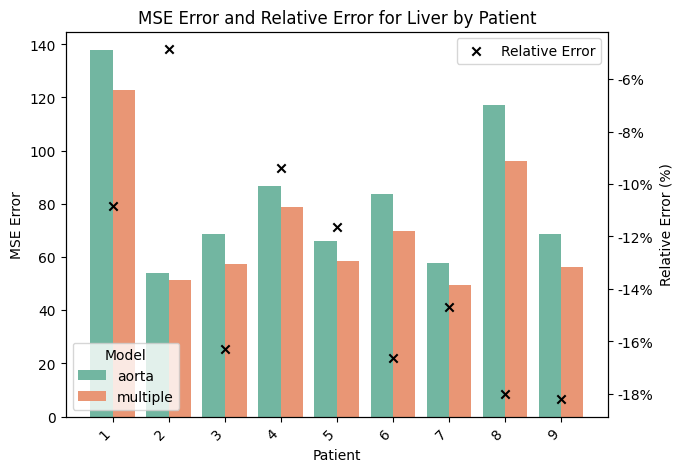}
    \includegraphics[width=0.4\linewidth]{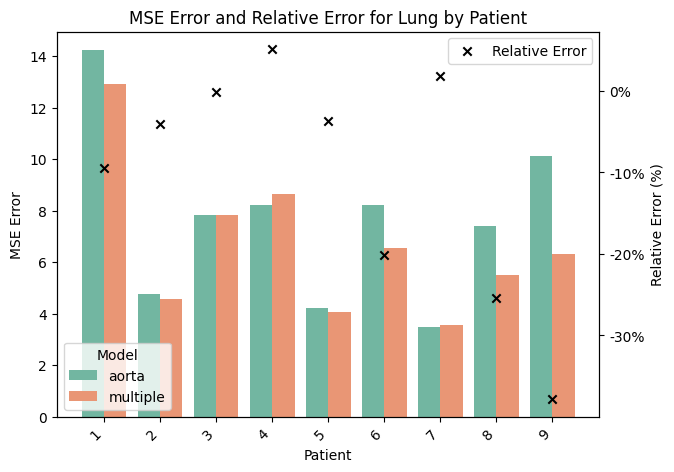}
    \caption{MSE values and relative error for liver and lung for all 9 patients.}
    \label{fig:MSE}
\end{figure}

\begin{figure}
    \centering
    \includegraphics[width=\textwidth]{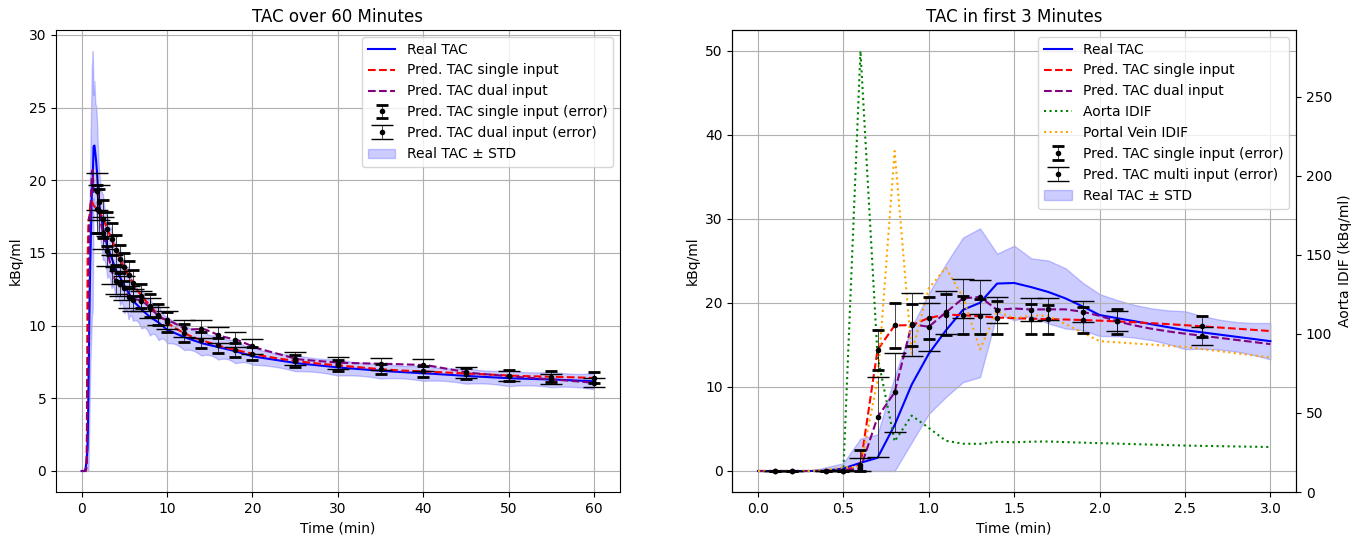}
    \caption{Predicted vs. measured TAC for liver for an exemplary patient.}
    \label{fig:TACliver}
\end{figure}

\begin{figure}
    \centering
    \includegraphics[width=\textwidth]{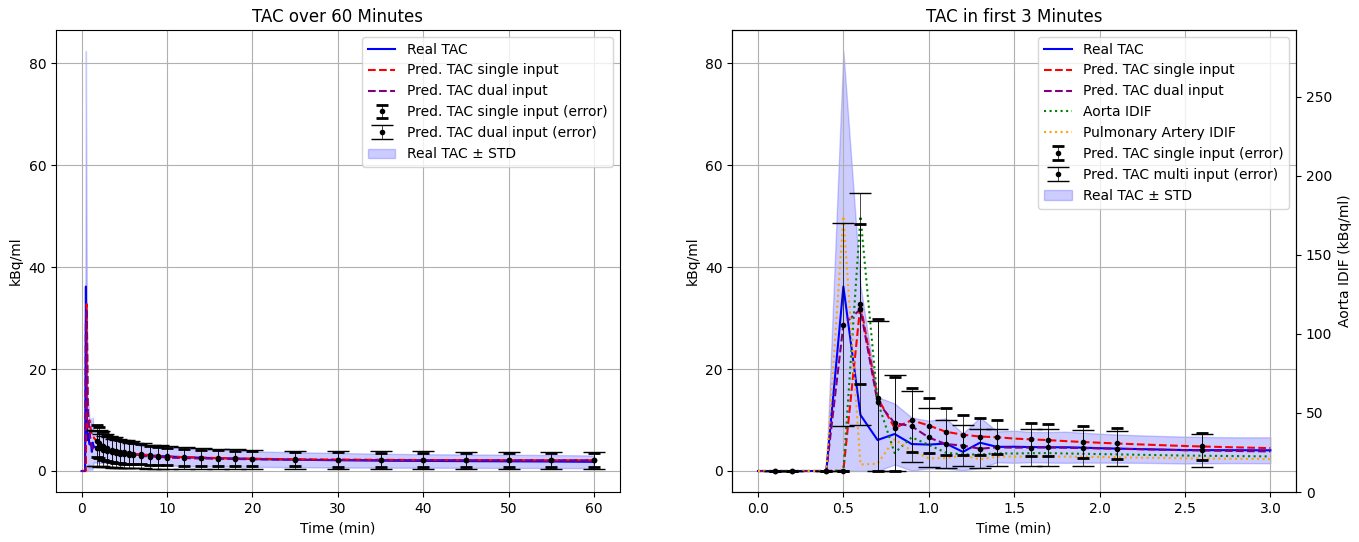}
    \caption{Predicted vs. measured TAC for lungs for an exemplary patient.}
    \label{fig:TAClung}
\end{figure}

\begin{figure}
    \centering
    \includegraphics[width=\textwidth]{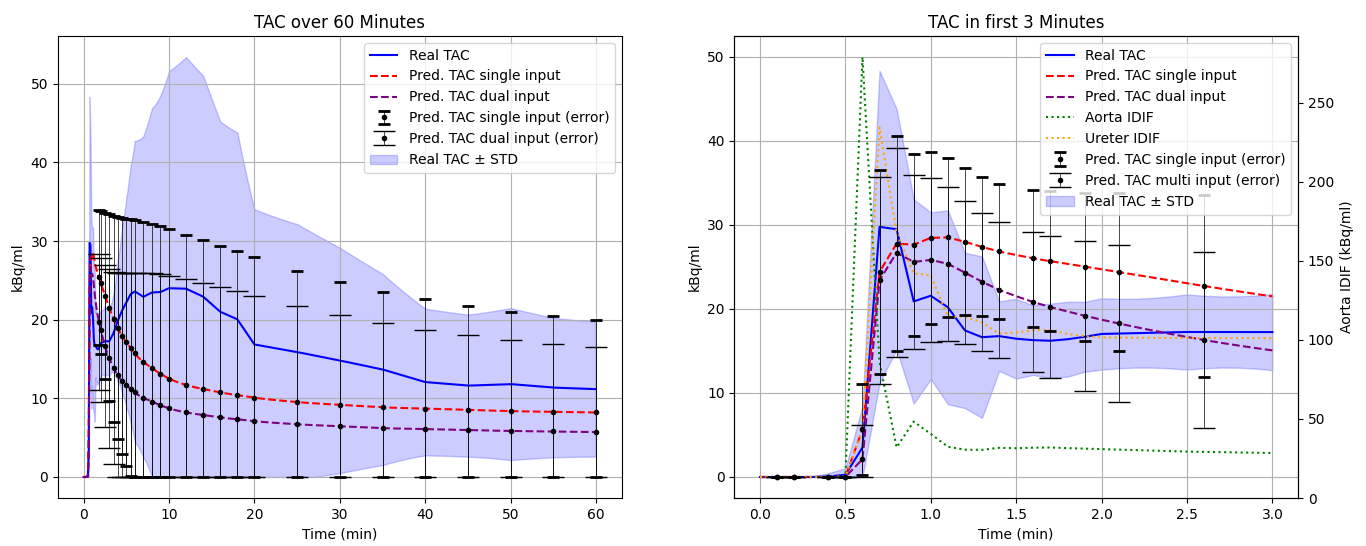}
    \caption{Predicted vs. measured TAC for kidneys for an exemplary patient.}
    \label{fig:TACkidney}
\end{figure}

We evaluated the MSE between measured and computed TACs for the liver, the lungs and the kidneys, both using only the IDIF of aorta (baseline) and multiple IDIFs depending on the organ of interest. We also quantified the relative change to the MSE and calculated its p-value using a Wilcoxon test (Tab. \ref{tab:MSE}). Additionally, we plotted the average MSE for each patient (Fig. \ref{fig:MSE}) and plotted the average TACs for all three organs for one exemplary patient (Figs. \ref{fig:TACliver}, \ref{fig:TAClung}, \ref{fig:TACkidney})

\begin{table}[h]
    \centering
    \caption{MSE (average $\pm$ standard deviation in $kBq/ml$), for selected organs using only IDIF of aorta or multiple IDIFs. * indicates a statistically significant difference for a 5\% significance level.}
    \begin{tabular}{lcccc}
    \hline
    Organ & MSE (aorta) & MSE (multiple) & Relative change & p-value \\
    \hline
    lung & 7.61 $\pm$ 3.31 & 6.66 $\pm$ 2.88 & -10.42\% $\pm$ 14.37\% & 0.07422\\
    liver & 82.28 $\pm$ 28.20 & 71.11 $\pm$ 24.49 & -13.39\% $\pm$ 4.52\% & 0.00391*\\
    kidney & 159.08 $\pm$ 61.16 & 158.86 $\pm$ 61.18 & -0.18\% $\pm$ 0.49\% & 0.42578\\
    \hline
    \vspace{0.2em}
    \end{tabular}
    
    \label{tab:MSE}
\end{table}

\subsection{Qualitative analysis full body analysis}

\begin{figure}
    \centering
    \includegraphics[width=\textwidth]{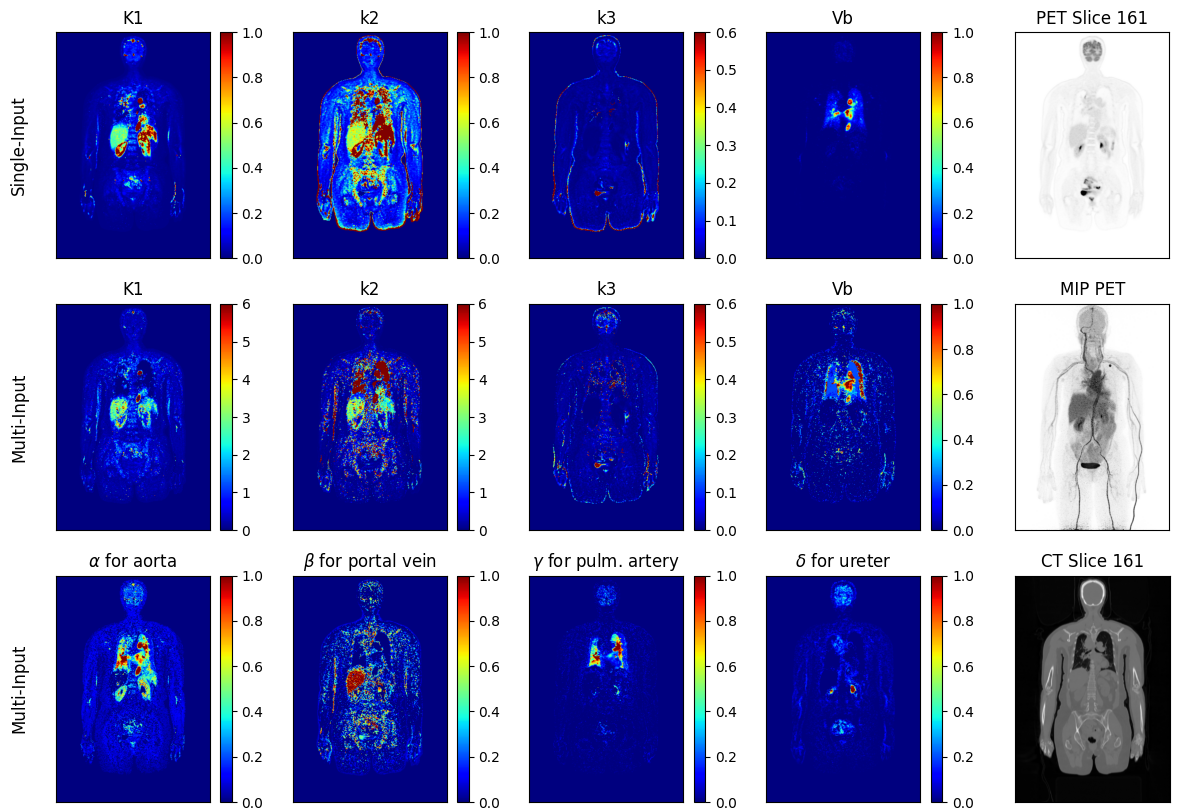}
    \caption{Coronal view of calculated parametric images only using the aorta (single input) (first row, first four columns) vs. images using multiple IDIFs (multiple input) (second row, first four columns), along with calculated $\alpha$, $\beta$, $\gamma$, and $\delta$ values (third row, first four columns), the corresponding PET and CT slice (last column, top and bottom), and an MIP of PET (last column, center) as reference.}
    \label{fig:coronal}
\end{figure}

\begin{figure}
    \centering
    \includegraphics[width=\textwidth]{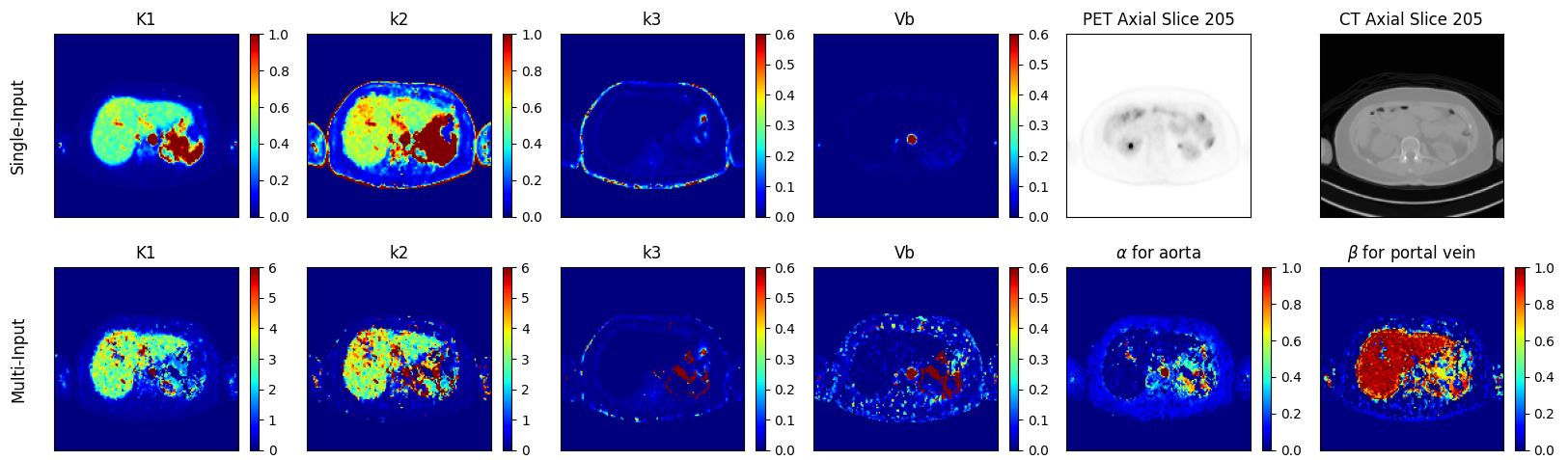}
    \caption{Axial view of calculated parametric images only using the aorta (single input) (first row, first four columns) vs. images using multiple IDIFs (multiple input) (second row, first four columns), along with calculated $\alpha$ and $\beta$, (second row, last two columns), and the corresponding PET and CT slice (first row, last two columns). The slice is taken across the liver.}
    \label{fig:liveraxial}
\end{figure}

\begin{figure}
    \centering
    \includegraphics[width=\textwidth]{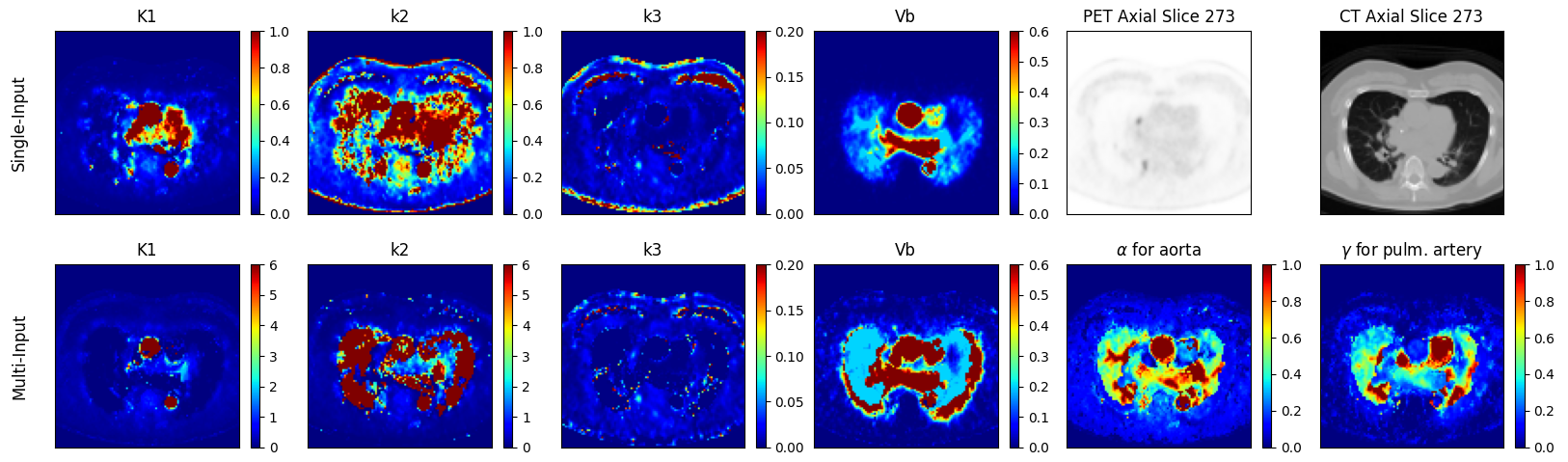}
    \caption{Axial view of calculated parametric images only using the aorta (single input) (first row, first four columns) vs. images using multiple IDIFs (multiple input) (second row, first four columns), along with calculated $\alpha$ and $\beta$, (second row, last two columns), and the corresponding PET and CT slice (first row, last two columns). The slice is taken across the lung.}
    \label{fig:lungaxial}
\end{figure}

To be able to grasp the impact of our method on the parametric images, we created coronal (Fig. \ref{fig:coronal}) and axial views across liver (Fig. \ref{fig:liveraxial}) and lung (Fig. \ref{fig:lungaxial}) for an exemplary patient.

\section{Discussion}

Accurate modelling of tracer kinetics requires accounting for the specific blood supply of each organ, as several receive tracer input from sources beyond the aorta, such as the portal vein, pulmonary arteries, and ureters \cite{brix2001quantification}. The descending aorta is commonly used as the IDIF \cite{volpi2023update}, but this approach assumes direct arterial supply to all tissues, overlooking the complexity of vascular pathways that influence tracer distribution. While IDIFs derived from smaller arteries provide an alternative \cite{tang2025assessment}, relying on a uniform aortic input may introduce inaccuracies in kinetic parameter estimation. To address this, our study incorporates anatomically accurate blood circulation sources by leveraging multi-organ segmentations from co-registered high-resolution CT scans, defining multiple IDIFs.

Our results demonstrate that using multiple IDIFs instead of a single aortic IDIF significantly improves model accuracy for certain organs. This effect is most pronounced for the liver, where the MSE was reduced by 13.39\% (p = 0.00391), indicating a substantial improvement in modelling accuracy.

For the lungs, we also observed an improvement, with an average MSE reduction of 10.42\%, though this was not statistically significant in our small patient cohort. In the case of the kidneys, the improvement was minimal, implying that either alternative blood supplies have a negligible impact on modelling or that a more complex approach is needed to accurately represent the excretion of FDG via the ureters. Furthermore, our approximation of ureter segmentation using the renal pelvis as a surrogate may not fully capture ureteral tracer kinetics. Future studies should explore direct ureter segmentation to enhance accuracy and better reflect renal excretion dynamics.

Overall, our findings suggest that incorporating multiple IDIFs from different vascular sources can enhance the accuracy of kinetic modelling. This approach is particularly promising for liver imaging, offering a viable alternative to conventional methods, while further optimizations may be needed for other organs.

\newpage 
\bibliography{Main.bib}

\end{document}